\documentclass[aps,prl,twocolumn,nofootinbib,longbibliography,superscriptaddress]{revtex4-1}
\usepackage{hyperref}
\usepackage{amsmath}
\usepackage{amssymb}
\usepackage{amsthm}
\usepackage{graphicx}
\usepackage{adjustbox}
\usepackage[dvipsnames]{xcolor}
\newcommand{\bi}{\begin{itemize}}
\newcommand{\ei}{\end{itemize}}
\newcommand{\be}{\begin{equation}}
\newcommand{\ee}{\end{equation}}
\newcommand{\ran}{\rangle}
\newcommand{\lan}{\langle}
\newcommand{\Tr}{\mathrm{Tr}}
\newcommand{\Hh}{\mathcal{H}}
\newcommand{\mO}{\mathcal{O}}
\newcommand{\wt}{\widetilde}
\newcommand{\bfig}{\begin{figure}\begin{center}}
\newcommand{\efig}{\end{center}\end{figure}}

\begin{document}
\title{Observer complementarity for black holes and holography}
\author{Netta Engelhardt}
\affiliation{Center for Theoretical Physics\\ Massachusetts Institute of Technology, Cambridge, MA 02139, USA}
\author{Elliott Gesteau}
\affiliation{Division of Physics, Mathematics, and Astronomy, California Institute of Technology,\\
Pasadena, CA 91125, USA}
\affiliation{Kavli Institute for Theoretical Physics, \\Santa Barbara, CA 93106, USA}
\author{Daniel Harlow}
\affiliation{Center for Theoretical Physics\\ Massachusetts Institute of Technology, Cambridge, MA 02139, USA}

\begin{abstract}
We present a mathematical formulation of black hole complementarity based on recent rules for including the observer in quantum cosmology.  We argue that this provides a self-consistent treatment of the interior of an evaporating black hole throughout its history, as well as the Antonini-Sasieta-Swingle-Rath configuration where a closed universe is entangled with a pair of AdS universes. 
\end{abstract}

\maketitle
\section{Introduction}
Complementarity -- the idea that the experiences of different observers can be incompatible so long as these observers cannot communicate -- was once thought to be crucial in the resolution of the black hole information problem~\cite{Susskind:1993if}.  This principle however was called into question in \cite{AMPS}, and so far it has not played a substantial role in the recent developments on the problem starting with \cite{Pen19, AEMM}.  For example one of the key ideas of complementarity was that a semiclassical description of an entire time slice of an evaporating black hole might not make sense, since no single observer has access to everything on such a slice \cite{Kiem:1995iy,Lowe:1995ac}.  The Page curve calculation of \cite{Pen19, AEMM} however does use such a description, and yet it leads to an answer (via the quantum extremal surface (QES) formula \cite{EngWal14}) that is consistent with unitary evaporation.  The interpretations of this calculation based on the path integral \cite{PenShe19,AlmHar19} and/or non-isometric codes \cite{AkeEng22} similarly have not involved complementarity in any obvious way.

In this paper we argue that a form of complementarity is nonetheless still necessary.  The reason is essentially as it was originally: at late times an external observer can verify that an outgoing mode near the horizon is entangled with the early radiation, while an infalling observer can verify that this mode is entangled with an interior partner, thus violating  monogamy of entanglement \cite{Mat09,Bra09V1}, \cite{AMPS}.  In the non-isometric code formalism of \cite{AkeEng22}, this apparent tension was resolved without complementarity: since the complexity of the first of these verifications is exponential in the entropy of the remaining black hole  \cite{HarHay13}, \cite{AkeEng22} proposed that the interior observer's experiences need not be semiclassical for such operations (in particular they might affect the state of the interior mode at spacelike separation).  It was recently noted however that this resolution does not work to describe the experience of an interior observer in a completely evaporated black hole \cite{HarUsa25,Bousso:2025udh}.  This is because distilling a purification of a late exterior Hawking mode from the rest of the radiation is easy once evaporation is complete.  So it seems that some form of complementarity is still needed to explain the emergence of the interior spacetime of a fully evaporated black hole.  

In order to realize complementarity as a mathematical formalism, we need a way for the laws of physics themselves to depend on which observer is performing measurements - otherwise there is no way for the same question to have different answers for different observers.  A technique for doing this has recently been proposed in the context of quantum gravity in a closed universe \cite{HarUsa25,AbdSte25} as a way of resolving the apparent consequence of standard holography that the Hilbert space of quantum gravity in a closed universe has dimension one \cite{AlmMah19a,PenShe19,McNVaf20,UsaWan24,EngGes25}.  In this article we will argue that applying this rule as a general principle for all observers leads to a theory of complementarity that can avoid the violation of entanglement monogamy without leading (as far as we can tell) to any inconsistency.  

\section{Review of emergent spacetime}
We begin by reviewing some features of the theory of emergent spacetime.    In \textit{any} situation where an ``effective description'' emerges from some ``fundamental'' quantum degrees of freedom, we can formulate this using a linear ``encoding'' map
\be
V:\Hh_{eff}\to \Hh_{fund}
\ee
from the effective Hilbert space to the fundamental Hilbert space.  For example in a solid, $\Hh_{eff}$ could be the Hilbert space of weakly-interacting phonons and $\Hh_{fund}$ the Hilbert space of the standard model of particle physics.  
For us $\Hh_{eff}$ is the Hilbert space of perturbative quantum gravity while $\Hh_{fund}$ is the Hilbert space of the fundamental holographic degrees of freedom (such as the boundary CFT in the AdS/CFT correspondence \cite{AlmDon14}).  The states in $\Hh_{fund}$ that are candidates for having an effective description are those of the form $V|\psi\ran$ for some $|\psi\ran\in \Hh_{eff}$; this description is good to the extent that we have
\be\label{isometry}
\lan \phi|V^\dagger V|\psi\ran\approx \lan \phi|\psi\ran.
\ee
If eq. \eqref{isometry} holds for all $|\psi\ran,|\phi\ran\in \Hh_{eff}$, then $V$ is approximately isometric.  In simple examples of emergence such as the phonon theory, $V$ is indeed an approximate isometry.  On the other hand the emergence of the spacetime inside of a black hole from its fundamental microstate degrees of freedom apparently requires $V$ to be highly non-isometric: the number of pertubative gravity states inside of a sufficiently old black hole is much larger than the number of microstates  \cite{Almheiri:2018xdw,PenShe19,MarMax20,Akers:2021fut,AkeEng22}.  \cite{AkeEng22} proposed the following framework: \eqref{isometry} should hold only for states whose complexity is subexponential in the entropy of the black hole and only up to corrections which are exponentially small in that entropy.  A model  encoding map with these properties is shown in figure \ref{evapcodefig}. 
\bfig
\includegraphics[height=4cm]{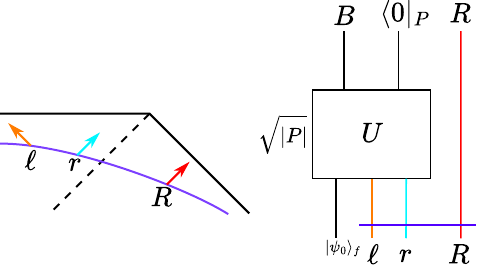}
\caption{A holographic code $V:\Hh_\ell\otimes \Hh_r\otimes \Hh_R\to \Hh_B\otimes \Hh_R$ for a partially evaporated black hole on a ``nice'' time slice, shaded purple.  The left and right interior modes $\ell$ and $r$ are acted upon with a unitary $U$, with some fixed auxiliary state $|\psi_0\ran_f$ included to restrict the set of possible states.  The unitary is followed by a projection $\lan 0|_P$ on a subset of the degrees of freedom and a rescaling by $\sqrt{|P|}$, where $|P|={\rm dim}\Hh_P$. The remaining system $B$ is the microstate Hilbert space of the black hole.  The encoding map acts trivially outside of the black hole, e.g. on the Hawking radiation $R$.}\label{evapcodefig}
\efig

\bfig
\includegraphics[height=4.8cm]{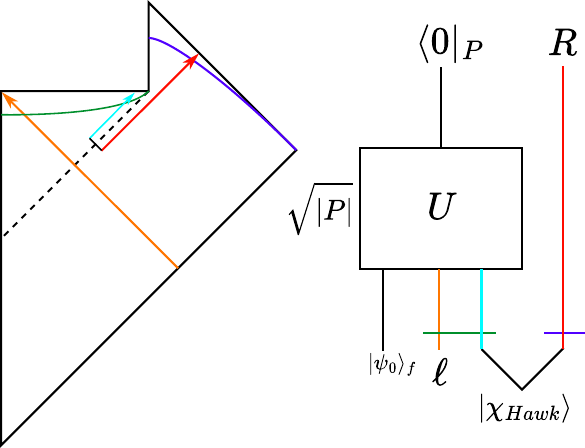}
\caption{A holographic code for a completely evaporated black hole, acting on a state created from an infalling shell (orange line). $|\chi_{Hawk}\ran$ is Hawking's state representing entanglement between interior and exterior outgoing modes (shaded light blue and red).  The baby universe part of the slice, shaded green, is subject to a rank-one projection, so the circuit on the right sends pure states of $\ell$ to pure states of $R$.}\label{fullevapfig}
\efig
Unfortunately the approach of \cite{AkeEng22} needs some modification before it can be applied to a fully evaporated black hole \cite{HarUsa25,Bousso:2025udh}.  See figure \ref{fullevapfig}: in the limit of complete evaporation the interior encoding map becomes a rank one projection.  This cannot be consistent with \eqref{isometry} for pretty much any states, essentially because errors suppressed by the black hole entropy become large since a completely evaporated black hole has no entropy.  This shows the close connection between the information problem and the statement that the fundamental Hilbert space of a closed universe is one-dimensional: in figure \ref{fullevapfig} the interior slice is essentially a closed universe, and if the code allowed any non-projected output of the unitary $U$ then pure states falling into the black hole wouldn't be mapped to pure states of the radiation $R$ (this completely-evaporated limit of the non-isometric code is similar to the Horowitz-Maldacena final state proposal \cite{HorMal03}, see \cite{Almheiri:2025ugo} for a recent discussion).

\section{Review of the observer rule}
To deal with the emergence of spacetime in a closed universe or the interior of a completely evaporated black hole, \cite{HarUsa25} proposed a way to treat the observer differently from the rest of the matter in the universe (a similar rule  introduced in \cite{AbdSte25} follows from the rule of \cite{HarUsa25} in the limit of infinite observer entropy). To state the rule we must consider the thorny question of what constitutes an observer.  In \cite{HarUsa25} three minimal assumptions were introduced:
\bi
\item[(1)] An observer has a number $S_{Ob}$ of quantum degrees of freedom, and they can only do experiments up to ambiguities/errors  of order $e^{-S_{Ob}}$.
\item[(2)] An observer should be approximately classical, meaning that they have a basis $|a\ran$ of approximate ``pointer states'' that are stable under interaction with the environment \cite{Zurek:2003zz}.  For example for Schrodinger's cat $|\mathrm{alive}\ran$ and $|\mathrm{dead}\ran$ are pointer states but $|\mathrm{alive}\ran+|\mathrm{dead}\ran$ is not.  
\item[(3)] Observers do not come in pure quantum states, i.e. in any state of interest the entanglement between the observer and their environment is of order $S_{Ob}$.
\ei
In the previous section we defined the inner product of encoded states to be $\lan \phi |V^\dagger V|\psi\ran$, but  this leads to trouble with a completely evaporated black hole or a closed universe.  The proposal of \cite{HarUsa25} is that instead of \eqref{isometry}
we should instead demand
\be\label{Obrule}
\Tr\left(V\mathcal{C}_{Ob}\left(|\psi\ran\lan \phi|\right)V^\dagger\right)\approx \lan \phi |\psi\ran,
\ee
where $\mathcal{C}_{Ob}$ is a \textit{quantum to classical channel} that acts on operators on the observer Hilbert space as
\be
\mathcal{C}_{Ob}\left(|a\ran\lan b|\right)=\delta_{ab}|a\ran\lan a|.
\ee
In other words $\mathcal{C}_{Ob}$ is a quantum channel that deletes the off-diagonal components of any operator on $Ob$ in the pointer basis.  This rule only makes sense on states $|\psi\ran$ and $|\phi\ran$ where an observer is present, but these are the only states where observations can be done.  The rule can be thought of as implementing a ``Heisenberg cut'', which in the Copenhagen interpretation of quantum mechanics is a dividing line between quantum and classical.  What was shown in \cite{HarUsa25} is that if we assume our observer obeys (1-3), then inside a closed universe or a completely evaporated black hole \eqref{Obrule} indeed holds up to corrections which are of order $e^{-S_{Ob}}$.  We will now argue that this rule introduces a form of complementarity.

\section{Applications}

\subsection{Antonini-Sasieta-Swingle-Rath geometry}
\bfig
\includegraphics[height=3.3cm]{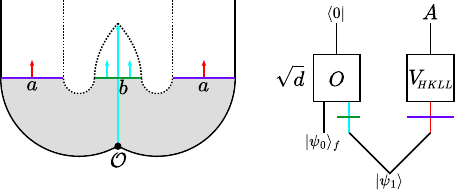}
\caption{A holographic encoding map for the ASSR geometry.  Left: the geometry.  The shaded region is the Euclidean preparation, the solid lines are asymptotic AdS boundaries, and the dotted lines are surfaces where the suppressed sphere shrinks to zero size.  The long blue line is the matter shell created by the boundary operator $\mO$ that sources the closed universe, for which we have indicated a Cauchy slice in green. The purple slices are Cauchy slices of the two AdS regions. Effective degrees of freedom in the AdS regions are denoted $a$, those in the closed universe are denoted $b$, and the boundary CFT degrees of freedom are denoted $A$.  The map is a rank-one projection on $b$ and the HKLL isometry on $a$, which maps the entangled state $|\psi_1\ran_{ab}$ to a pure state of $A$.}\label{arcodefig}
\efig
Before taking on black holes, we will first show how the observer rule introduces complementarity in a simpler situation introduced in \cite{AntSas23} and advocated in \cite{AntRat24} as a particularly nice laboratory for understanding closed universes using AdS/CFT.  The idea is to modify the thermofield double state of two holographic CFTs to include a spherically-symmetric heavy operator $\mO$ midway along the Euclidean boundary.  This introduces a shell of dust in the dual geometry whose mass density is of order $1/G$.  For sufficiently long Euclidean boundaries, this leads to a dominant bulk geometry which in Lorentzian signature consists of two approximately AdS universes containing gases of matter particles that are jointly entangled with some matter particles in a closed universe whose size in AdS units is large.  The amount of matter entanglement is large but of order $G^0$, and the closed universe experiences a big crunch in a time which is of order the AdS time.  See the left side of figure \ref{arcodefig} for an illustration of this geometry, which we will call the ASSR geometry.

The holographic encoding map $V:\Hh_a\otimes \Hh_b\to \Hh_A$ for the ASSR geometry is quite simple: on the two AdS regions we use the HKLL encoding isometry \cite{HamKab05}, while the closed universe is hit by a rank one projection taken to be a $d$-dimensional orthogonal transformation $O$ on the closed universe together with some fixed state $|\psi_0\ran_f$ followed by a projection onto a simple state $\lan 0|$ \cite{HarUsa25}. See the right side of figure \ref{arcodefig} for an illustration. Orthogonality of the matrix $O$ follows from the gauging of CRT symmetry (which implies that the Hilbert space of a closed universe is real) \cite{Harlow:2023hjb}.  This code gives a simple interpretation of a puzzling feature of this state  pointed out in \cite{AntRat24}. Namely the bulk state $|\psi_1\ran\in \mathcal{H}_a\otimes\mathcal{H}_b$ prepared by the Euclidean path integral has the feature that there is another bulk state $|\psi_2\ran\in \mathcal{H}_a$ on the AdS regions alone such that
\be\label{AReq}
V|\psi_1\ran\propto V_{HKLL}|\psi_2\ran.
\ee
The state $|\psi_2\ran$ describes two nearly-AdS geometries, each with a gas of matter such that the two gases are together in an entangled pure state, with no baby universe anywhere.\footnote{The argument that such a $|\psi_2\ran_a$ exists proceeds by showing that that $V|\psi_1\ran_{ab}$ is mostly supported on states of energy which is $O(G^0)$, and thus must be in the image of $V_{HKLL}$ \cite{AntRat24,EngGes25}.}  From the code we explicitly have
\be\label{psi2}
|\psi_2\ran=\frac{\lan 0|O|\psi_1\ran}{\sqrt{\lan \psi_1|O^T|0\ran\lan 0|O|\psi_1\ran}}.
\ee
Equation \eqref{AReq} is interesting because it seems to contradict the idea that each CFT state should have a unique bulk interpretation \cite{AntRat24}.  Does a bulk observer think the state is $|\psi_1\ran$ or $|\psi_2\ran$?  This was left ambiguous in \cite{AntRat24}, but in \cite{EngGes25} it was emphasized that the minimal AdS/CFT dictionary tells us that at least for a boundary observer the answer must be $|\psi_2\ran$.  This is because  ${\rm exp}(-S_{2}(V\psi_1 V^\dagger)_A)$, which can be interpreted as the expectation value of a simple bulk operator $\mathcal{S}_{AdS}$ on two copies of the system ($\mathcal{S}_{AdS}$ exchanges the AdS gases between the two copies), unambiguously tells us that the state of the gas in the two AdS regions is pure.  We now discuss the situation for bulk observers using the observer rule.

\bfig
\includegraphics[height=4cm]{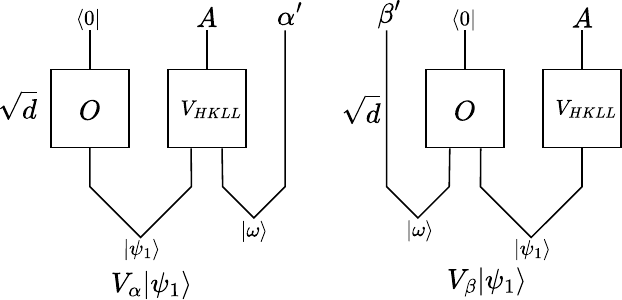}
\caption{Observer codes for the ASSR geometry (suppressing $|\psi_0\ran_f$).  Cloning the observer out of the system to implement the  quantum-to-classical channel creates an entangled state $|\omega\ran$ of the observer, their environment, and their clone.}\label{ARobsfig}
\efig
We consider two bulk observers in the ASSR geometry: 
\bi
\item An observer $\alpha$ in one of the AdS regions.
\item An observer $\beta$ in the baby universe.
\ei 
The easiest way to implement the quantum-to-classical channel in \eqref{Obrule} is to modify the holographic map to clone the observer in their pointer basis to some external system $\alpha',\beta'$ \cite{HarUsa25}.  The modified holographic maps $V_\alpha$ and $V_\beta$ implementing the observer rule for our two observers are shown in figure \ref{ARobsfig}. Using these maps, we find\footnote{The $V_\alpha$ code is a little subtle, as it doesn't do a good job preserving the inner product from the closed universe part of the state.  This is a feature and not a bug, as it tells us that the closed universe doesn't exist for the $\alpha$ observer, but to get a code that does work we need to re-interpret $V_\alpha$ a bit to ``excise'' the closed universe.  We explain this further in appendix A.}
\begin{align}\label{ASSRa}
\lan \wt{\mathcal{S}}_{AdS}\ran_\alpha&\approx 1\\
\lan\wt{\mathcal{S}}_{AdS} \ran_\beta&\approx \max\left[e^{-S_2(\psi_{1,a})},e^{-S_2(\omega_{\beta})}\right].\label{ASSRb}
\end{align}
Here $\lan \cdot \ran_{\alpha,\beta}$ indicate the encoded expectation values in the state $|\psi_1\ran\otimes |\psi_1\ran$ for the $\alpha$ and $\beta$ codes, and $\wt{\mathcal{S}}_{AdS}=V_{HKLL}\mathcal{S}_{AdS}V_{HKLL}^\dagger$ is the encoded swap operator.  See appendix A for more on these calculations.  Thus we see that an observer in the AdS regions thinks that the state of the AdS gases is pure, in agreement with a boundary observer, but an observer in the closed universe finds a smaller expectation value for $\mathcal{S}_{AdS}$.  Indeed by assumption (3) we have $S_2(\omega_\beta)\sim S_{\beta}$, so \eqref{ASSRb} is consistent with the state being $|\psi_1\ran$ up to errors which are of order $e^{-S_\beta}$, which by assumption (1) is the best $\beta$ can ask for.  This is our first instance of complementarity: $\alpha$ thinks the bulk state is $|\psi_2\ran$ but $\beta$ thinks it is $|\psi_1\ran$. We emphasize however that there is no way for $\alpha$ and $\beta$ to argue about this, as they live in disconnected components of the universe.

\subsection{Evaporating black hole}
We now turn to the evaporating black hole codes shown in figures \ref{evapcodefig} and \ref{fullevapfig}.  If we introduce an exterior observer $\alpha$ and an infalling observer $\beta$ who falls in, then on average (over $U$ in the Haar measure) we have (see appendix A) 
\begin{align}\label{bhswapa}
\lan \mathcal{S}_R\ran_\alpha&\approx\max\left[e^{-S_2(\chi_{Hawk},R)},\frac{1}{|B|}\right]\\\label{bhswapb}
\lan \mathcal{S}_R\ran_\beta&\approx\max\left[e^{-S_2(\chi_{Hawk},R)},\frac{e^{-S_2(\omega_\beta)}}{|B|}\right].
\end{align}
The first of these is the standard ``Page curve'' result for an evaporating black hole: at early times the Renyi entropy matches Hawking's calculation while at late times it matches the coarse-grained entropy of the remaining black hole.  In particular in the completely-evaporated limit the expectation value of swap goes to one since the radiation state is pure.  The $\beta$ result however is more novel: while the black hole is still big (meaning $\log |B| \gg S_2(\omega_\beta)$) then we essentially get the same result as for the $\alpha$ code, but in the completely evaporated limit $|B|\to 1$ we can instead interpret the result as saying that $\beta$ thinks the swap expectation value agrees with Hawking's result $e^{-S_2(\chi_{Hawk,R})}$ up to errors  of order $e^{-S_\beta}$.  In other words $\alpha$ and $\beta$ do not agree about the final state of the radiation: $\alpha$ thinks it is pure, while $\beta$ thinks it is consistent with Hawking's result.  This again is a form of complementarity, and just as in the ASSR situation they cannot communicate this disagreement.  In fact we can make this inability more quantitative: say that before the black hole has finished evaporating an $\alpha$ observer changes her mind and decides to jump in.  In order for her to fit, her entropy must obey $S_\alpha \ll \log |B|$, in which case her swap result \eqref{bhswapa} is again consistent with Hawking's result $e^{-S_2(\chi_{Hawk,R})}$ up to errors of order $e^{-S_\alpha}$. 

Two other commonly-considered tasks in this context are the distillation of a purification of a late Hawking mode from the early Hawking radiation and the reconstruction of an interior unitary operator $W_{\ell r}$ as a unitary operator on the Hawking radiation.  Information-theoretically both of these are controlled by decoupling theorems, with the result being that we can do both if (see appendix A) 
\begin{align}\nonumber
S_2(\chi_{Hawk,R})&\gg 1+\log |B|\qquad\phantom{+S_\beta} \,\,(\alpha \, \mathrm{code})\\
S_2(\chi_{Hawk,R})&\gg 1+\log |B|+S_\beta\qquad (\beta \, \mathrm{code}).\label{decoupling}
\end{align}
It is interesting to consider the complexity of these tasks.  In the partially evaporated case the distillation/reconstruction must proceed without access to the degrees of freedom in the remaining black hole, in which case by \cite{HarHay13,BroGha19} they are expected to have complexity which is exponential in the black hole entropy $\log |B|$.  The situation is more subtle when the black hole is completely evaporated. For the code shown in figure \ref{fullevapfig} these tasks are still exponentially complex post-evaporation, i) because the Haar random unitary $U$ is exponentially complex and ii) because ``undoing'' the post-selection $\lan 0|_P$ requires a Grover amplification whose complexity goes like $\sqrt{|P|}$ \cite{BroGha19}.  Both of these are unrealistic artifacts however, as for a true black hole the unitary S-matrix which takes the initial state of $\ell$ to the final state on $R$ should have only polynomial complexity in the entropy of the initial black hole.  A more refined code that respects this was introduced in \cite{AkeEng22}, and we show in appendix B that in the $\alpha$ version of this code both distillation and reconstruction are easy while in the $\beta$ version reconstruction is easy but distillation has complexity which is exponential in $S_\beta$.  The easy reconstruction in the $\beta$ code may seem concerning, but in appendix B we show that whenever this reconstruction is easy there is a large gravitational backreaction which contaminates the $\beta$ observer and invalidates Hawking's framework.  Thus we see that in all cases $\alpha$ sees results that are consistent with unitary evaporation while $\beta$ sees results (for observables of subexponential complexity in $S_\beta$) that are consistent with Hawking's picture whenever the latter makes sense.

\section{Towards a general formulation}

We have modeled the experiences of two observers, $\alpha$ and $\beta$, of entropy $S_\alpha$ and $S_\beta$, in the cases of the baby universe geometry of ASSR and an evaporating black hole, using the observer rule of \cite{HarUsa25}. Table \ref{recaptab} summarizes our findings for the various tasks that we have considered.
\begin{table}
\begin{adjustbox}{width=\columnwidth,center}
\begin{tabular}{lll}
\toprule
$\langle\widetilde{\mathcal{S}}\rangle$ & ASSR & Black Hole\\
$\alpha$ & 1 & $1/\vert B\vert$\\
$\beta$  & $e^{-S_2(\psi_{1,a})}+\mathcal{O}(e^{-S_{\beta}})$ & $e^{-S_2(\chi_{Hawk,R})}+\mathcal{O}(e^{-S_\beta})$ \\
Semiclassical & $e^{-S_2(\psi_{1,a})}$ & $e^{-S_2(\chi_{Hawk,R})}$\\
\toprule
AMPS Distillation & Partially Evaporated & Fully Evaporated\\
$\alpha$ & EXP($\log |B|$)$\sim {\rm EXP}(1/G_{N})$ & POLY($S_2(\chi_{Hawk,R})$)\\
$\beta$  & $\gtrsim$EXP($S_\beta$) &$\gtrsim$ EXP($S_\beta$) \\
Semiclassical & $\infty$ & $\infty$\\
\toprule
Left Mover Reconst. & Partially Evaporated & Fully Evaporated\\
$\alpha$ & EXP($\log |B|$) $\sim {\rm EXP}(1/G_{N})$ & POLY($S_2(\chi_{Hawk,R})$)\\
$\beta$  &$\gtrsim$ EXP($S_\beta$) & POLY($S_2(\chi_{Hawk,R})$) \\
Semiclassical & $\infty$ & x \\
\end{tabular}
\end{adjustbox}
\caption{Predictions for $\alpha$ and $\beta$ compared to the semiclassical picture for the expectation value of swap, AMPS distillation, and left-moving reconstruction.} 
\label{recaptab}
\end{table}
We observe (pun intended) that the three following principles remain true throughout our calculations:
\begin{itemize}
\item[(1)] $\alpha$ finds answers consistent with the unitarity of the holographic description up to errors of order $e^{-S_\alpha}$.
\item[(2)] $\beta$ finds answers for observables subexponential in $S_\beta$ that are consistent with the semiclassical picture when the latter is valid up to errors of order $e^{-S_\beta}$.
\item[(3)] Observers $\alpha$ and $\beta$ agree when they can causally communicate, up errors of order $e^{-\min(S_\alpha,S_\beta)}$. 
\end{itemize}
We find it plausible that these can be promoted to general principles for complementarity in holographic systems.

\section*{Acknowledgments}
It is a pleasure to thank S. Aaronson, R. Bousso, and G. Penington for valuable discussions.    The work of NE is supported in part by the Department of Energy under Early Career Award DE-SC0021886, by the Heising-Simons Foundation under grant no. 2023-4430,  and by the Templeton Foundation via the Black Hole Initiative.
The work of EG was supported in part by the the Todd Alworth Larson fellowship, the Heising-Simons Foundation, the Simons Foundation, and grants no. NSF PHY-2309135 to the Kavli Institute for Theoretical Physics (KITP). The work of DH is supported by the Packard Foundation, the US Department of Energy under grants DE-SC0012567 and DE-SC0020360, and the MIT Department of Physics.

\appendix
\section{Appendix A: Calculations using the observer rule}\label{calcapp}
\bfig
\includegraphics[height=3.8cm]{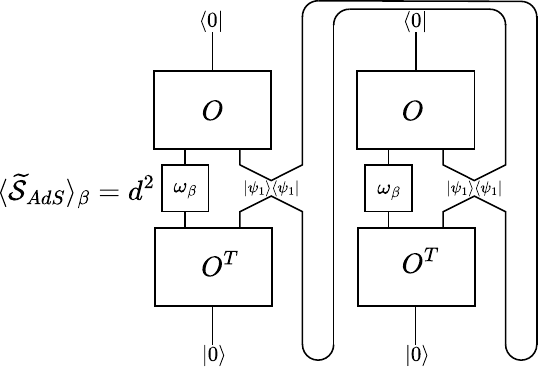}
\caption{Computing the expectation value of the encoded swap operator $\wt{\mathcal{S}}_{AdS}$ in the $\beta$ code.}\label{betaswapfig}
\efig
In this appendix we give some more details on calculations from the main text, starting with the swap results \eqref{ASSRa},\eqref{ASSRb}.  The $V_\beta$ code is a bit easier to think about, the index contractions are shown in figure \ref{betaswapfig}. Using the orthogonal integration technology reviewed in appendix A of \cite{HarUsa25} we have (see also equation 4.5 of \cite{HarUsa25})
{\small \begin{align}\nonumber
\int dO \lan \wt{\mathcal{S}}_{AdS}\ran_\beta&=\int dO \lan \psi_1|V_\beta^\dagger \otimes \lan\psi_1|V_\beta^\dagger \wt{\mathcal{S}}_{AdS} V_\beta|\psi_1\ran\otimes V_\beta|\psi_1\ran\\\nonumber
&=\frac{d}{d+2}\Big[\Tr\left(\psi_{1,a}^2\right)+\Tr\left(\omega_\beta^2\right)\\
&\phantom{\frac{d}{d+2}\Big[}+\Tr\left(\psi_{1,a}\psi_{1,a}^T\right)\Tr\left(\omega_\beta\omega_\beta^T\right)\Big].
\end{align}}
\hspace{-.15cm}Here $\psi_{1,a}$ is the reduction of $|\psi_1\ran$ to $a$ and $\omega_\beta$ is the reduction of $|\omega\ran$ to the observer $\beta$.  We are interested in the large $d$ limit, in which case we can ignore the prefactor, and the third term is subleading to the first two since $|\psi_1\ran$ and $|\omega\ran$ are not close to being product states (had we used a unitary matrix instead of an orthogonal matrix this third term would be absent). This completes the derivation of \eqref{ASSRb}.  

Turning now to the $V_\alpha$ code, the complication which arises is that the inner product on the baby universe is not well preserved.  For example we have
{\small\be
\int dO \lan \psi_1 |V_\alpha^\dagger V_\alpha|\psi_1\ran^2=\frac{d}{d+2}\left[1+\Tr\left(\psi_{1,a}^2\right)+\Tr\left(\psi_{1,a}\psi_{1,a}^T\right)\right],
\ee}
\hspace{-.1cm}so the norm of the encoded state is only perserved up to fluctuations which are exponentially suppressed by $S_2(\psi_{1,a})$.  This entropy is not so large in the ASSR state, so we can easily have $S_\alpha\gg S_2(\psi_{1,a})$ in which case the $V_\alpha$ code is not consistent with the emergence of a semiclassical baby universe.  These fluctuations directly affect the expectation value of the swap operator:
{\small\begin{align}\nonumber
\int dO \lan \wt{\mathcal{S}}_{AdS}\ran_\alpha&=\int dO \lan \psi_1|V_\alpha^\dagger \otimes \lan\psi_1|V_\alpha^\dagger \wt{\mathcal{S}}_{AdS} V_\alpha|\psi_1\ran\otimes V_\alpha|\psi_1\ran\\
&=\frac{d}{d+2}\Big[1+\Tr\left(\psi_{1,a}^2\right)+\Tr\left(\psi_{1,a}\psi_{1,a}^T\right)\Big].
\end{align}}
\hspace{-.1cm}The first term here is the one we want to reproduce \eqref{ASSRa}, while the second two are inherited from the fluctuations of the norm of $V_\alpha|\psi_1\ran$.  The way to deal with these fluctuations is apparent from considering equations \eqref{AReq} and \eqref{psi2}: what \eqref{AReq} is telling us is that the $\alpha$ observer should really view $|\psi_2\ran$ as the input to the code rather than $|\psi_1\ran$.  The closed universe part of the code converts $|\psi_1\ran$ to $|\psi_2\ran$, but this succeeds only up to an overall normalization which is restored by hand in equation \eqref{psi2}.  We get a better code if we simply remove the closed universe, leading to a modified code $\hat{V}_\alpha$ shown in figure \ref{alpha2codefig}.  In this code the encoding map is just $V_{HKLL}$, and we simply get $\lan\wt{\mathcal{S}}_{AdS}\ran_\alpha=1$ on the nose since $V_{HKLL}$ is an isometry.  We emphasize that in this calculation the swap operator acts on the gas leg going into $V_{HKLL}$ but \textit{not} on the observer leg: it is the former that we want to be pure.

\bfig
\includegraphics[height=3cm]{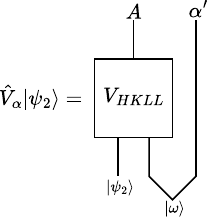}
\caption{Modifying the $\alpha$ observer's code to excise the closed universe.}\label{alpha2codefig}
\efig
The code $\hat{V}_\alpha$ is essentially the same as the code $V_\alpha$, except that we have ``excised'' the baby universe region which is causally inaccessible to $\alpha$.  You may ask why we have not similarly excised the AdS regions in the $\beta$ code.  Indeed we could have, and arguably it would be more in keeping with the principle of complementarity.  The reasons we didn't are i) it isn't necessary to get a good code and ii) had we done so we would need to use a different observable than $\mathcal{S}_{AdS}$ to diagnose the entanglement in $\psi_1$.  We think it is better to err on the side of a ``more global'' code, at least until this leads to a problem.  When it does, the global picture has failed to emerge and then we need to excise until we get a code that works.

The swap calculation for a black hole is similar to that for the ASSR setup, except that now there are B legs to worry about.  We'll work with an encoded state
\be
\rho_{BR\alpha',\beta'}=V_{\alpha,\beta}\left(|\psi\ran\lan \psi|_\ell \otimes |\chi_{Hawk}\ran\lan \chi_{Hawk}|_{rR}\right)V_{\alpha,\beta}^\dagger.
\ee
In the $\beta$ code the result (see equation 4.4 of \cite{AkeEng22}) is that
{\small\begin{align}\nonumber
\int dU \Tr\left(\rho_R^2\right)=\frac{|B|^2|P|^2}{|B|^2|P|^2-1}\Bigg[&e^{-S_2(\chi_{Hawk,R})}\left(1-\frac{1}{|P||B|^2}\right)\\
&+\frac{1}{|B|}e^{-S_2(\omega_\beta)}\left(1-\frac{1}{|P|}\right)\Bigg],
\end{align}}
\hspace{-.1cm}which agrees with \eqref{bhswapa}, \eqref{bhswapb} in the limit of large $|P|$ (to get the $\alpha$ code we send $S_2(\omega_\beta)\to 0$).  The decoupling calculations for distillation and reconstruction are quite similar to this one, for example for reconstruction what we want to show is that the family of encoded states
{\small\be
\rho_{BR\alpha',\beta'}(W)=V_{\alpha,\beta}\left(W_{\ell r}|\psi\ran\lan \psi|_\ell \otimes |\chi_{Hawk}\ran\lan \chi_{Hawk}|_{rR}W_{\ell r}^\dagger\right)V_{\alpha,\beta}^\dagger
\ee}
\hspace{-.15cm}obeys $||\rho_{B\alpha',\beta'}(W)-\rho_{B\alpha',\beta'}(W')||_1\ll 1$ for all $W$ and $W'$ (see theorem 5.2 from \cite{AkeEng22}).  From equation 5.27 of \cite{AkeEng22} we have
\be
||\rho_{B\alpha',\beta'}(W)-\rho_{B\alpha',\beta'}(W')||_1\leq 2 e^{-(S_2(\chi_{Hawk,R})-\log |B|-S_{\beta})/2},
\ee
which confirms \eqref{decoupling}.

\section{Appendix B: Backreaction and reconstruction for the evaporated black hole}
\begin{figure}
\centering
\includegraphics[scale=0.5]{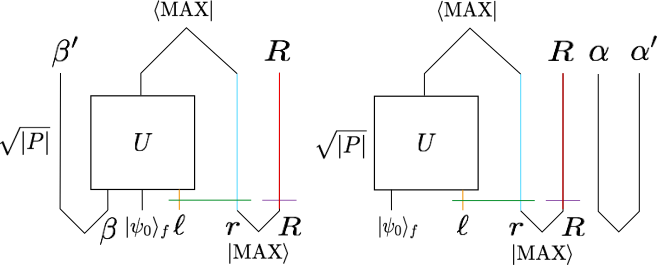}
\caption{A refined code for a completely evaporated black hole \cite{AkeEng22}, supplemented with the observer rules for $\alpha$ and $\beta$.   Here $U$ is a random unitary of polynomial depth, as opposed to a Haar random unitary, and $|\chi_{Hawk}\ran$ has been simplified to be maximally entangled.  By ``straightening'' the $R$ leg we see $U$ is in fact just the black hole S-matrix sending $\Hh_\ell$ to $\Hh_R$.}
\label{bettercodefig1}
\end{figure}
The black hole codes in figures \ref{evapcodefig}, \ref{fullevapfig} serve well to address questions that do not involve details of past dynamics of a fully evaporated black hole.  We expect this to be the case for information-theoretic questions such as when distillation and reconstruction are possible, but not for dynamical questions such as difficulty of implementation of these operations.  In particular any dynamical operation involving the late stages of an evaporating black hole must deal with Planckian curvatures in the final stages of the evaporation; these can lead to effects that invalidate the semiclassical Hawking picture.  To better model such dynamical effects, a more refined code was introduced in \cite{AkeEng22}.  The $\alpha$ and $\beta$ versions of this code for a completely evaporated black hole are shown in figure \ref{bettercodefig1}.

The results \eqref{bhswapa}, \eqref{bhswapb}, \eqref{decoupling} for swap and decoupling are similar for this code and the simpler codes in figures \ref{evapcodefig}, \ref{fullevapfig}, but the distillation and reconstruction complexities are different in the fully evaporated case. In either of the observer codes in figure \ref{bettercodefig1}, to act on the ingoing $\ell$ leg with a unitary operator $W_\ell$ by doing something in the radiation we can simply use the polynomial operation 
\begin{align}
\widetilde{W}_R=UW_\ell U^\dagger,
\end{align}
since the $r$ legs can be ``straightened out". This corresponds to reversing time evolution for the black hole, whose S-matrix is of polynomial complexity in the initial black hole entropy, and it works in both the $\alpha$ and $\beta$ codes.  If we more generally wish to reconstruct an operator $W_{\ell r}$ which acts on both left and right movers in the interior this simple algorithm fails, but Grover amplification does furnish a reconstruction whose complexity is exponential in the number of $r$ modes which are affected by $W_{\ell r}$ \cite{EngPen21b},\cite{AkeEng22} so this reconstruction is still not difficult if only a few $r$ modes are affected.  On the other hand the complexity of distilling a purification of a single Hawking mode depends on whether we use the $\alpha$ or $\beta$ code in figure \ref{bettercodefig1}: in the $\alpha$ code we have access to the full final state so by using Grover amplification it is easy to distill a purification.  In the $\beta$ code however there is still entanglement with $\beta'$, so by the results of \cite{HarHay13,BroGha19} the distillation is exponential in $S_\beta$.

\begin{figure}
    \centering
    \includegraphics[width=0.8\linewidth]{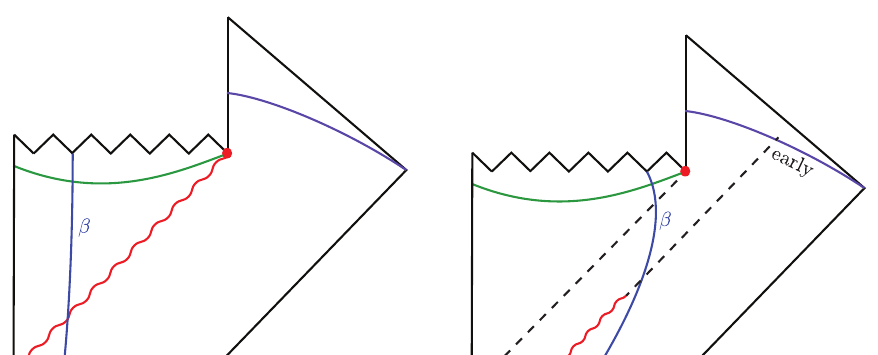}
    \caption{Left panel: Any reconstruction task involving the late radiation necessarily evolves back through the Planckian curvatures near evaporation point (red dot), likely leading to a singular horizon that invalidates the history of the $\beta$ observer.  Right panel: reconstructing only on the early radiation still results in a singularity upon backwards time evolution, but by jumping in after the early radiation came out, $\beta$ can evade this singularity and remain semiclassical.}
    \label{fig:appendB1}
\end{figure}
The ``easy'' reconstruction of $W_\ell$ in the $\beta$ code seems to be in some tension with principle 2 above.  It is important to remember however that this reconstruction must act on \textit{all} of the radiation. Any description of its effect on the semiclassical picture thus requires evolving all the radiation backwards in time. Since the black hole is completely evaporated, this evolution takes some of the qubits of the radiation through the region of large backreaction near the singularity, leading to a large backreaction which affects the past of the observer and invalidates the Hawking description of their experiences.\footnote{We could try to circumvent this large backreaction by doing the reconstruction on just the early radiation.  However, in this case reconstruction for $\beta$ becomes exponentially complex in $S_{\beta}$: this setup with its null initial data is now simply the Python's lunch of the partially evaporated black hole~\cite{BroGha19, EngPen21b}, so we are again consistent with principle 2.}  See figure~\ref{fig:appendB1}. The code still does give an answer for what they should see, but it arises from a timefold: the reconstruction evolves the black hole S-matrix backwards and then changes the initial state so then when we evolve forward again the $\beta$ observer now sees something different inside.  The apparent ability of this simple reconstruction operation to send information outside the lightcone is analogous to the situation with timefolds near the ordinary AdS vacuum \cite{HeeMar12} (in particular if we try to interpret acting with a time-evolved boundary operator using effective field theory in AdS then we get a similar shockwave to that shown in figure \ref{fig:appendB1}).
\bibliography{all}

\begin{thebibliography}{34}%
\makeatletter
\providecommand \@ifxundefined [1]{%
 \@ifx{#1\undefined}
}%
\providecommand \@ifnum [1]{%
 \ifnum #1\expandafter \@firstoftwo
 \else \expandafter \@secondoftwo
 \fi
}%
\providecommand \@ifx [1]{%
 \ifx #1\expandafter \@firstoftwo
 \else \expandafter \@secondoftwo
 \fi
}%
\providecommand \natexlab [1]{#1}%
\providecommand \enquote  [1]{``#1''}%
\providecommand \bibnamefont  [1]{#1}%
\providecommand \bibfnamefont [1]{#1}%
\providecommand \citenamefont [1]{#1}%
\providecommand \href@noop [0]{\@secondoftwo}%
\providecommand \href [0]{\begingroup \@sanitize@url \@href}%
\providecommand \@href[1]{\@@startlink{#1}\@@href}%
\providecommand \@@href[1]{\endgroup#1\@@endlink}%
\providecommand \@sanitize@url [0]{\catcode `\\12\catcode `\$12\catcode
  `\&12\catcode `\#12\catcode `\^12\catcode `\_12\catcode `\%12\relax}%
\providecommand \@@startlink[1]{}%
\providecommand \@@endlink[0]{}%
\providecommand \url  [0]{\begingroup\@sanitize@url \@url }%
\providecommand \@url [1]{\endgroup\@href {#1}{\urlprefix }}%
\providecommand \urlprefix  [0]{URL }%
\providecommand \Eprint [0]{\href }%
\providecommand \doibase [0]{http://dx.doi.org/}%
\providecommand \selectlanguage [0]{\@gobble}%
\providecommand \bibinfo  [0]{\@secondoftwo}%
\providecommand \bibfield  [0]{\@secondoftwo}%
\providecommand \translation [1]{[#1]}%
\providecommand \BibitemOpen [0]{}%
\providecommand \bibitemStop [0]{}%
\providecommand \bibitemNoStop [0]{.\EOS\space}%
\providecommand \EOS [0]{\spacefactor3000\relax}%
\providecommand \BibitemShut  [1]{\csname bibitem#1\endcsname}%
\let\auto@bib@innerbib\@empty
\bibitem [{\citenamefont {Susskind}\ \emph {et~al.}(1993)\citenamefont
  {Susskind}, \citenamefont {Thorlacius},\ and\ \citenamefont
  {Uglum}}]{Susskind:1993if}%
  \BibitemOpen
  \bibfield  {author} {\bibinfo {author} {\bibfnamefont {Leonard}\ \bibnamefont
  {Susskind}}, \bibinfo {author} {\bibfnamefont {Larus}\ \bibnamefont
  {Thorlacius}}, \ and\ \bibinfo {author} {\bibfnamefont {John}\ \bibnamefont
  {Uglum}},\ }\bibfield  {title} {\enquote {\bibinfo {title} {{The Stretched
  horizon and black hole complementarity}},}\ }\href {\doibase
  10.1103/PhysRevD.48.3743} {\bibfield  {journal} {\bibinfo  {journal} {Phys.
  Rev. D}\ }\textbf {\bibinfo {volume} {48}},\ \bibinfo {pages} {3743--3761}
  (\bibinfo {year} {1993})},\ \Eprint {http://arxiv.org/abs/hep-th/9306069}
  {arXiv:hep-th/9306069} \BibitemShut {NoStop}%
\bibitem [{\citenamefont {Almheiri}\ \emph {et~al.}(2013)\citenamefont
  {Almheiri}, \citenamefont {Marolf}, \citenamefont {Polchinski},\ and\
  \citenamefont {Sully}}]{AMPS}%
  \BibitemOpen
  \bibfield  {author} {\bibinfo {author} {\bibfnamefont {Ahmed}\ \bibnamefont
  {Almheiri}}, \bibinfo {author} {\bibfnamefont {Donald}\ \bibnamefont
  {Marolf}}, \bibinfo {author} {\bibfnamefont {Joseph}\ \bibnamefont
  {Polchinski}}, \ and\ \bibinfo {author} {\bibfnamefont {James}\ \bibnamefont
  {Sully}},\ }\bibfield  {title} {\enquote {\bibinfo {title} {{Black Holes:
  Complementarity or Firewalls?}}}\ }\href {\doibase 10.1007/JHEP02(2013)062}
  {\bibfield  {journal} {\bibinfo  {journal} {JHEP}\ }\textbf {\bibinfo
  {volume} {02}},\ \bibinfo {pages} {062} (\bibinfo {year} {2013})},\ \Eprint
  {http://arxiv.org/abs/1207.3123} {arXiv:1207.3123 [hep-th]} \BibitemShut
  {NoStop}%
\bibitem [{\citenamefont {Penington}(2020)}]{Pen19}%
  \BibitemOpen
  \bibfield  {author} {\bibinfo {author} {\bibfnamefont {Geoffrey}\
  \bibnamefont {Penington}},\ }\bibfield  {title} {\enquote {\bibinfo {title}
  {{Entanglement Wedge Reconstruction and the Information Paradox}},}\ }\href
  {\doibase 10.1007/JHEP09(2020)002} {\bibfield  {journal} {\bibinfo  {journal}
  {JHEP}\ }\textbf {\bibinfo {volume} {09}},\ \bibinfo {pages} {002} (\bibinfo
  {year} {2020})},\ \Eprint {http://arxiv.org/abs/1905.08255} {arXiv:1905.08255
  [hep-th]} \BibitemShut {NoStop}%
\bibitem [{\citenamefont {Almheiri}\ \emph {et~al.}(2019)\citenamefont
  {Almheiri}, \citenamefont {Engelhardt}, \citenamefont {Marolf},\ and\
  \citenamefont {Maxfield}}]{AEMM}%
  \BibitemOpen
  \bibfield  {author} {\bibinfo {author} {\bibfnamefont {Ahmed}\ \bibnamefont
  {Almheiri}}, \bibinfo {author} {\bibfnamefont {Netta}\ \bibnamefont
  {Engelhardt}}, \bibinfo {author} {\bibfnamefont {Donald}\ \bibnamefont
  {Marolf}}, \ and\ \bibinfo {author} {\bibfnamefont {Henry}\ \bibnamefont
  {Maxfield}},\ }\bibfield  {title} {\enquote {\bibinfo {title} {{The entropy
  of bulk quantum fields and the entanglement wedge of an evaporating black
  hole}},}\ }\href {\doibase 10.1007/JHEP12(2019)063} {\bibfield  {journal}
  {\bibinfo  {journal} {JHEP}\ }\textbf {\bibinfo {volume} {12}},\ \bibinfo
  {pages} {063} (\bibinfo {year} {2019})},\ \Eprint
  {http://arxiv.org/abs/1905.08762} {arXiv:1905.08762 [hep-th]} \BibitemShut
  {NoStop}%
\bibitem [{\citenamefont {Kiem}\ \emph {et~al.}(1995)\citenamefont {Kiem},
  \citenamefont {Verlinde},\ and\ \citenamefont {Verlinde}}]{Kiem:1995iy}%
  \BibitemOpen
  \bibfield  {author} {\bibinfo {author} {\bibfnamefont {Youngjai}\
  \bibnamefont {Kiem}}, \bibinfo {author} {\bibfnamefont {Herman~L.}\
  \bibnamefont {Verlinde}}, \ and\ \bibinfo {author} {\bibfnamefont {Erik~P.}\
  \bibnamefont {Verlinde}},\ }\bibfield  {title} {\enquote {\bibinfo {title}
  {{Black hole horizons and complementarity}},}\ }\href {\doibase
  10.1103/PhysRevD.52.7053} {\bibfield  {journal} {\bibinfo  {journal} {Phys.
  Rev. D}\ }\textbf {\bibinfo {volume} {52}},\ \bibinfo {pages} {7053--7065}
  (\bibinfo {year} {1995})},\ \Eprint {http://arxiv.org/abs/hep-th/9502074}
  {arXiv:hep-th/9502074} \BibitemShut {NoStop}%
\bibitem [{\citenamefont {Lowe}\ \emph {et~al.}(1995)\citenamefont {Lowe},
  \citenamefont {Polchinski}, \citenamefont {Susskind}, \citenamefont
  {Thorlacius},\ and\ \citenamefont {Uglum}}]{Lowe:1995ac}%
  \BibitemOpen
  \bibfield  {author} {\bibinfo {author} {\bibfnamefont {David~A.}\
  \bibnamefont {Lowe}}, \bibinfo {author} {\bibfnamefont {Joseph}\ \bibnamefont
  {Polchinski}}, \bibinfo {author} {\bibfnamefont {Leonard}\ \bibnamefont
  {Susskind}}, \bibinfo {author} {\bibfnamefont {Larus}\ \bibnamefont
  {Thorlacius}}, \ and\ \bibinfo {author} {\bibfnamefont {John}\ \bibnamefont
  {Uglum}},\ }\bibfield  {title} {\enquote {\bibinfo {title} {{Black hole
  complementarity versus locality}},}\ }\href {\doibase
  10.1103/PhysRevD.52.6997} {\bibfield  {journal} {\bibinfo  {journal} {Phys.
  Rev. D}\ }\textbf {\bibinfo {volume} {52}},\ \bibinfo {pages} {6997--7010}
  (\bibinfo {year} {1995})},\ \Eprint {http://arxiv.org/abs/hep-th/9506138}
  {arXiv:hep-th/9506138} \BibitemShut {NoStop}%
\bibitem [{\citenamefont {Engelhardt}\ and\ \citenamefont
  {Wall}(2015)}]{EngWal14}%
  \BibitemOpen
  \bibfield  {author} {\bibinfo {author} {\bibfnamefont {Netta}\ \bibnamefont
  {Engelhardt}}\ and\ \bibinfo {author} {\bibfnamefont {Aron~C.}\ \bibnamefont
  {Wall}},\ }\bibfield  {title} {\enquote {\bibinfo {title} {{Quantum Extremal
  Surfaces: Holographic Entanglement Entropy beyond the Classical Regime}},}\
  }\href {\doibase 10.1007/JHEP01(2015)073} {\bibfield  {journal} {\bibinfo
  {journal} {JHEP}\ }\textbf {\bibinfo {volume} {01}},\ \bibinfo {pages} {073}
  (\bibinfo {year} {2015})},\ \Eprint {http://arxiv.org/abs/1408.3203}
  {arXiv:1408.3203 [hep-th]} \BibitemShut {NoStop}%
\bibitem [{\citenamefont {Penington}\ \emph {et~al.}(2022)\citenamefont
  {Penington}, \citenamefont {Shenker}, \citenamefont {Stanford},\ and\
  \citenamefont {Yang}}]{PenShe19}%
  \BibitemOpen
  \bibfield  {author} {\bibinfo {author} {\bibfnamefont {Geoff}\ \bibnamefont
  {Penington}}, \bibinfo {author} {\bibfnamefont {Stephen~H.}\ \bibnamefont
  {Shenker}}, \bibinfo {author} {\bibfnamefont {Douglas}\ \bibnamefont
  {Stanford}}, \ and\ \bibinfo {author} {\bibfnamefont {Zhenbin}\ \bibnamefont
  {Yang}},\ }\bibfield  {title} {\enquote {\bibinfo {title} {{Replica wormholes
  and the black hole interior}},}\ }\href {\doibase 10.1007/JHEP03(2022)205}
  {\bibfield  {journal} {\bibinfo  {journal} {JHEP}\ }\textbf {\bibinfo
  {volume} {03}},\ \bibinfo {pages} {205} (\bibinfo {year} {2022})},\ \Eprint
  {http://arxiv.org/abs/1911.11977} {arXiv:1911.11977 [hep-th]} \BibitemShut
  {NoStop}%
\bibitem [{\citenamefont {Almheiri}\ \emph
  {et~al.}(2020{\natexlab{a}})\citenamefont {Almheiri}, \citenamefont
  {Hartman}, \citenamefont {Maldacena}, \citenamefont {Shaghoulian},\ and\
  \citenamefont {Tajdini}}]{AlmHar19}%
  \BibitemOpen
  \bibfield  {author} {\bibinfo {author} {\bibfnamefont {Ahmed}\ \bibnamefont
  {Almheiri}}, \bibinfo {author} {\bibfnamefont {Thomas}\ \bibnamefont
  {Hartman}}, \bibinfo {author} {\bibfnamefont {Juan}\ \bibnamefont
  {Maldacena}}, \bibinfo {author} {\bibfnamefont {Edgar}\ \bibnamefont
  {Shaghoulian}}, \ and\ \bibinfo {author} {\bibfnamefont {Amirhossein}\
  \bibnamefont {Tajdini}},\ }\bibfield  {title} {\enquote {\bibinfo {title}
  {{Replica Wormholes and the Entropy of Hawking Radiation}},}\ }\href
  {\doibase 10.1007/JHEP05(2020)013} {\bibfield  {journal} {\bibinfo  {journal}
  {JHEP}\ }\textbf {\bibinfo {volume} {05}},\ \bibinfo {pages} {013} (\bibinfo
  {year} {2020}{\natexlab{a}})},\ \Eprint {http://arxiv.org/abs/1911.12333}
  {arXiv:1911.12333 [hep-th]} \BibitemShut {NoStop}%
\bibitem [{\citenamefont {Akers}\ \emph {et~al.}(2024)\citenamefont {Akers},
  \citenamefont {Engelhardt}, \citenamefont {Harlow}, \citenamefont
  {Penington},\ and\ \citenamefont {Vardhan}}]{AkeEng22}%
  \BibitemOpen
  \bibfield  {author} {\bibinfo {author} {\bibfnamefont {Chris}\ \bibnamefont
  {Akers}}, \bibinfo {author} {\bibfnamefont {Netta}\ \bibnamefont
  {Engelhardt}}, \bibinfo {author} {\bibfnamefont {Daniel}\ \bibnamefont
  {Harlow}}, \bibinfo {author} {\bibfnamefont {Geoff}\ \bibnamefont
  {Penington}}, \ and\ \bibinfo {author} {\bibfnamefont {Shreya}\ \bibnamefont
  {Vardhan}},\ }\bibfield  {title} {\enquote {\bibinfo {title} {{The black hole
  interior from non-isometric codes and complexity}},}\ }\href {\doibase
  10.1007/JHEP06(2024)155} {\bibfield  {journal} {\bibinfo  {journal} {JHEP}\
  }\textbf {\bibinfo {volume} {06}},\ \bibinfo {pages} {155} (\bibinfo {year}
  {2024})},\ \Eprint {http://arxiv.org/abs/2207.06536} {arXiv:2207.06536
  [hep-th]} \BibitemShut {NoStop}%
\bibitem [{\citenamefont {Mathur}(2009)}]{Mat09}%
  \BibitemOpen
  \bibfield  {author} {\bibinfo {author} {\bibfnamefont {Samir~D.}\
  \bibnamefont {Mathur}},\ }\bibfield  {title} {\enquote {\bibinfo {title}
  {{The Information paradox: A Pedagogical introduction}},}\ }\href {\doibase
  10.1088/0264-9381/26/22/224001} {\bibfield  {journal} {\bibinfo  {journal}
  {Class.Quant.Grav.}\ }\textbf {\bibinfo {volume} {26}},\ \bibinfo {pages}
  {224001} (\bibinfo {year} {2009})},\ \Eprint {http://arxiv.org/abs/0909.1038}
  {arXiv:0909.1038 [hep-th]} \BibitemShut {NoStop}%
\bibitem [{\citenamefont {Braunstein}(2009)}]{Bra09V1}%
  \BibitemOpen
  \bibfield  {author} {\bibinfo {author} {\bibfnamefont {Samuel~L.}\
  \bibnamefont {Braunstein}},\ }\bibfield  {title} {\enquote {\bibinfo {title}
  {{Black hole entropy as entropy of entanglement, or it's curtains for the
  equivalence principle}},}\ }\href@noop {} {\  (\bibinfo {year} {2009})},\
  \Eprint {http://arxiv.org/abs/0907.1190v1} {arXiv:0907.1190v1 [quant-ph]}
  \BibitemShut {NoStop}%
\bibitem [{\citenamefont {Harlow}\ and\ \citenamefont
  {Hayden}(2013)}]{HarHay13}%
  \BibitemOpen
  \bibfield  {author} {\bibinfo {author} {\bibfnamefont {Daniel}\ \bibnamefont
  {Harlow}}\ and\ \bibinfo {author} {\bibfnamefont {Patrick}\ \bibnamefont
  {Hayden}},\ }\bibfield  {title} {\enquote {\bibinfo {title} {{Quantum
  Computation vs. Firewalls}},}\ }\href {\doibase 10.1007/JHEP06(2013)085}
  {\bibfield  {journal} {\bibinfo  {journal} {JHEP}\ }\textbf {\bibinfo
  {volume} {1306}},\ \bibinfo {pages} {085} (\bibinfo {year} {2013})},\ \Eprint
  {http://arxiv.org/abs/1301.4504} {arXiv:1301.4504 [hep-th]} \BibitemShut
  {NoStop}%
\bibitem [{\citenamefont {Harlow}\ \emph {et~al.}(2025)\citenamefont {Harlow},
  \citenamefont {Usatyuk},\ and\ \citenamefont {Zhao}}]{HarUsa25}%
  \BibitemOpen
  \bibfield  {author} {\bibinfo {author} {\bibfnamefont {Daniel}\ \bibnamefont
  {Harlow}}, \bibinfo {author} {\bibfnamefont {Mykhaylo}\ \bibnamefont
  {Usatyuk}}, \ and\ \bibinfo {author} {\bibfnamefont {Ying}\ \bibnamefont
  {Zhao}},\ }\bibfield  {title} {\enquote {\bibinfo {title} {{Quantum mechanics
  and observers for gravity in a closed universe}},}\ }\href@noop {} {\
  (\bibinfo {year} {2025})},\ \Eprint {http://arxiv.org/abs/2501.02359}
  {arXiv:2501.02359 [hep-th]} \BibitemShut {NoStop}%
\bibitem [{\citenamefont {Bousso}(2025)}]{Bousso:2025udh}%
  \BibitemOpen
  \bibfield  {author} {\bibinfo {author} {\bibfnamefont {Raphael}\ \bibnamefont
  {Bousso}},\ }\bibfield  {title} {\enquote {\bibinfo {title} {{Firewalls From
  General Covariance}},}\ }\href@noop {} {\  (\bibinfo {year} {2025})},\
  \Eprint {http://arxiv.org/abs/2502.08724} {arXiv:2502.08724 [hep-th]}
  \BibitemShut {NoStop}%
\bibitem [{\citenamefont {Abdalla}\ \emph {et~al.}(2025)\citenamefont
  {Abdalla}, \citenamefont {Antonini}, \citenamefont {Iliesiu},\ and\
  \citenamefont {Levine}}]{AbdSte25}%
  \BibitemOpen
  \bibfield  {author} {\bibinfo {author} {\bibfnamefont {Ahmed~I.}\
  \bibnamefont {Abdalla}}, \bibinfo {author} {\bibfnamefont {Stefano}\
  \bibnamefont {Antonini}}, \bibinfo {author} {\bibfnamefont {Luca~V.}\
  \bibnamefont {Iliesiu}}, \ and\ \bibinfo {author} {\bibfnamefont {Adam}\
  \bibnamefont {Levine}},\ }\bibfield  {title} {\enquote {\bibinfo {title}
  {{The gravitational path integral from an observer's point of view}},}\
  }\href@noop {} {\  (\bibinfo {year} {2025})},\ \Eprint
  {http://arxiv.org/abs/2501.02632} {arXiv:2501.02632 [hep-th]} \BibitemShut
  {NoStop}%
\bibitem [{\citenamefont {Almheiri}\ \emph
  {et~al.}(2020{\natexlab{b}})\citenamefont {Almheiri}, \citenamefont
  {Mahajan}, \citenamefont {Maldacena},\ and\ \citenamefont
  {Zhao}}]{AlmMah19a}%
  \BibitemOpen
  \bibfield  {author} {\bibinfo {author} {\bibfnamefont {Ahmed}\ \bibnamefont
  {Almheiri}}, \bibinfo {author} {\bibfnamefont {Raghu}\ \bibnamefont
  {Mahajan}}, \bibinfo {author} {\bibfnamefont {Juan}\ \bibnamefont
  {Maldacena}}, \ and\ \bibinfo {author} {\bibfnamefont {Ying}\ \bibnamefont
  {Zhao}},\ }\bibfield  {title} {\enquote {\bibinfo {title} {{The Page curve of
  Hawking radiation from semiclassical geometry}},}\ }\href {\doibase
  10.1007/JHEP03(2020)149} {\bibfield  {journal} {\bibinfo  {journal} {JHEP}\
  }\textbf {\bibinfo {volume} {03}},\ \bibinfo {pages} {149} (\bibinfo {year}
  {2020}{\natexlab{b}})},\ \Eprint {http://arxiv.org/abs/1908.10996}
  {arXiv:1908.10996 [hep-th]} \BibitemShut {NoStop}%
\bibitem [{\citenamefont {McNamara}\ and\ \citenamefont
  {Vafa}(2020)}]{McNVaf20}%
  \BibitemOpen
  \bibfield  {author} {\bibinfo {author} {\bibfnamefont {Jacob}\ \bibnamefont
  {McNamara}}\ and\ \bibinfo {author} {\bibfnamefont {Cumrun}\ \bibnamefont
  {Vafa}},\ }\bibfield  {title} {\enquote {\bibinfo {title} {{Baby Universes,
  Holography, and the Swampland}},}\ }\href@noop {} {\  (\bibinfo {year}
  {2020})},\ \Eprint {http://arxiv.org/abs/2004.06738} {arXiv:2004.06738
  [hep-th]} \BibitemShut {NoStop}%
\bibitem [{\citenamefont {Usatyuk}\ \emph {et~al.}(2024)\citenamefont
  {Usatyuk}, \citenamefont {Wang},\ and\ \citenamefont {Zhao}}]{UsaWan24}%
  \BibitemOpen
  \bibfield  {author} {\bibinfo {author} {\bibfnamefont {Mykhaylo}\
  \bibnamefont {Usatyuk}}, \bibinfo {author} {\bibfnamefont {Zi-Yue}\
  \bibnamefont {Wang}}, \ and\ \bibinfo {author} {\bibfnamefont {Ying}\
  \bibnamefont {Zhao}},\ }\bibfield  {title} {\enquote {\bibinfo {title}
  {{Closed universes in two dimensional gravity}},}\ }\href {\doibase
  10.21468/SciPostPhys.17.2.051} {\bibfield  {journal} {\bibinfo  {journal}
  {SciPost Phys.}\ }\textbf {\bibinfo {volume} {17}},\ \bibinfo {pages} {051}
  (\bibinfo {year} {2024})},\ \Eprint {http://arxiv.org/abs/2402.00098}
  {arXiv:2402.00098 [hep-th]} \BibitemShut {NoStop}%
\bibitem [{\citenamefont {Engelhardt}\ and\ \citenamefont
  {Gesteau}(2025)}]{EngGes25}%
  \BibitemOpen
  \bibfield  {author} {\bibinfo {author} {\bibfnamefont {Netta}\ \bibnamefont
  {Engelhardt}}\ and\ \bibinfo {author} {\bibfnamefont {Elliott}\ \bibnamefont
  {Gesteau}},\ }\bibfield  {title} {\enquote {\bibinfo {title} {{Further
  Evidence Against a Semiclassical Baby Universe in AdS/CFT}},}\ }\href@noop {}
  {\  (\bibinfo {year} {2025})},\ \Eprint {http://arxiv.org/abs/2504.14586}
  {arXiv:2504.14586 [hep-th]} \BibitemShut {NoStop}%
\bibitem [{\citenamefont {Almheiri}\ \emph {et~al.}(2015)\citenamefont
  {Almheiri}, \citenamefont {Dong},\ and\ \citenamefont {Harlow}}]{AlmDon14}%
  \BibitemOpen
  \bibfield  {author} {\bibinfo {author} {\bibfnamefont {Ahmed}\ \bibnamefont
  {Almheiri}}, \bibinfo {author} {\bibfnamefont {Xi}~\bibnamefont {Dong}}, \
  and\ \bibinfo {author} {\bibfnamefont {Daniel}\ \bibnamefont {Harlow}},\
  }\bibfield  {title} {\enquote {\bibinfo {title} {{Bulk Locality and Quantum
  Error Correction in AdS/CFT}},}\ }\href {\doibase 10.1007/JHEP04(2015)163}
  {\bibfield  {journal} {\bibinfo  {journal} {JHEP}\ }\textbf {\bibinfo
  {volume} {04}},\ \bibinfo {pages} {163} (\bibinfo {year} {2015})},\ \Eprint
  {http://arxiv.org/abs/1411.7041} {arXiv:1411.7041 [hep-th]} \BibitemShut
  {NoStop}%
\bibitem [{\citenamefont {Almheiri}(2018)}]{Almheiri:2018xdw}%
  \BibitemOpen
  \bibfield  {author} {\bibinfo {author} {\bibfnamefont {Ahmed}\ \bibnamefont
  {Almheiri}},\ }\bibfield  {title} {\enquote {\bibinfo {title} {{Holographic
  Quantum Error Correction and the Projected Black Hole Interior}},}\
  }\href@noop {} {\  (\bibinfo {year} {2018})},\ \Eprint
  {http://arxiv.org/abs/1810.02055} {arXiv:1810.02055 [hep-th]} \BibitemShut
  {NoStop}%
\bibitem [{\citenamefont {Marolf}\ and\ \citenamefont
  {Maxfield}(2020)}]{MarMax20}%
  \BibitemOpen
  \bibfield  {author} {\bibinfo {author} {\bibfnamefont {Donald}\ \bibnamefont
  {Marolf}}\ and\ \bibinfo {author} {\bibfnamefont {Henry}\ \bibnamefont
  {Maxfield}},\ }\bibfield  {title} {\enquote {\bibinfo {title} {{Transcending
  the ensemble: baby universes, spacetime wormholes, and the order and disorder
  of black hole information}},}\ }\href {\doibase 10.1007/JHEP08(2020)044}
  {\bibfield  {journal} {\bibinfo  {journal} {JHEP}\ }\textbf {\bibinfo
  {volume} {08}},\ \bibinfo {pages} {044} (\bibinfo {year} {2020})},\ \Eprint
  {http://arxiv.org/abs/2002.08950} {arXiv:2002.08950 [hep-th]} \BibitemShut
  {NoStop}%
\bibitem [{\citenamefont {Akers}\ and\ \citenamefont
  {Penington}(2022)}]{Akers:2021fut}%
  \BibitemOpen
  \bibfield  {author} {\bibinfo {author} {\bibfnamefont {Chris}\ \bibnamefont
  {Akers}}\ and\ \bibinfo {author} {\bibfnamefont {Geoff}\ \bibnamefont
  {Penington}},\ }\bibfield  {title} {\enquote {\bibinfo {title} {{Quantum
  minimal surfaces from quantum error correction}},}\ }\href {\doibase
  10.21468/SciPostPhys.12.5.157} {\bibfield  {journal} {\bibinfo  {journal}
  {SciPost Phys.}\ }\textbf {\bibinfo {volume} {12}},\ \bibinfo {pages} {157}
  (\bibinfo {year} {2022})},\ \Eprint {http://arxiv.org/abs/2109.14618}
  {arXiv:2109.14618 [hep-th]} \BibitemShut {NoStop}%
\bibitem [{\citenamefont {Horowitz}\ and\ \citenamefont
  {Maldacena}(2004)}]{HorMal03}%
  \BibitemOpen
  \bibfield  {author} {\bibinfo {author} {\bibfnamefont {Gary~T.}\ \bibnamefont
  {Horowitz}}\ and\ \bibinfo {author} {\bibfnamefont {Juan~Martin}\
  \bibnamefont {Maldacena}},\ }\bibfield  {title} {\enquote {\bibinfo {title}
  {{The Black Hole Final State}},}\ }\href {\doibase
  10.1088/1126-6708/2004/02/008} {\bibfield  {journal} {\bibinfo  {journal}
  {JHEP}\ }\textbf {\bibinfo {volume} {0402}},\ \bibinfo {pages} {008}
  (\bibinfo {year} {2004})},\ \Eprint {http://arxiv.org/abs/hep-th/0310281}
  {arXiv:hep-th/0310281 [hep-th]} \BibitemShut {NoStop}%
\bibitem [{\citenamefont {Almheiri}(2025)}]{Almheiri:2025ugo}%
  \BibitemOpen
  \bibfield  {author} {\bibinfo {author} {\bibfnamefont {Ahmed}\ \bibnamefont
  {Almheiri}},\ }\bibfield  {title} {\enquote {\bibinfo {title} {{Measurements
  $\mathit{with}$ probabilities in the final state proposal}},}\ }\href@noop {}
  {\  (\bibinfo {year} {2025})},\ \Eprint {http://arxiv.org/abs/2505.23664}
  {arXiv:2505.23664 [hep-th]} \BibitemShut {NoStop}%
\bibitem [{\citenamefont {Zurek}(2003)}]{Zurek:2003zz}%
  \BibitemOpen
  \bibfield  {author} {\bibinfo {author} {\bibfnamefont {Wojciech~Hubert}\
  \bibnamefont {Zurek}},\ }\bibfield  {title} {\enquote {\bibinfo {title}
  {{Decoherence, einselection, and the quantum origins of the classical}},}\
  }\href {\doibase 10.1103/RevModPhys.75.715} {\bibfield  {journal} {\bibinfo
  {journal} {Rev. Mod. Phys.}\ }\textbf {\bibinfo {volume} {75}},\ \bibinfo
  {pages} {715--775} (\bibinfo {year} {2003})},\ \Eprint
  {http://arxiv.org/abs/quant-ph/0105127} {arXiv:quant-ph/0105127} \BibitemShut
  {NoStop}%
\bibitem [{\citenamefont {Antonini}\ \emph {et~al.}(2023)\citenamefont
  {Antonini}, \citenamefont {Sasieta},\ and\ \citenamefont
  {Swingle}}]{AntSas23}%
  \BibitemOpen
  \bibfield  {author} {\bibinfo {author} {\bibfnamefont {Stefano}\ \bibnamefont
  {Antonini}}, \bibinfo {author} {\bibfnamefont {Martin}\ \bibnamefont
  {Sasieta}}, \ and\ \bibinfo {author} {\bibfnamefont {Brian}\ \bibnamefont
  {Swingle}},\ }\bibfield  {title} {\enquote {\bibinfo {title} {{Cosmology from
  random entanglement}},}\ }\href {\doibase 10.1007/JHEP11(2023)188} {\bibfield
   {journal} {\bibinfo  {journal} {JHEP}\ }\textbf {\bibinfo {volume} {11}},\
  \bibinfo {pages} {188} (\bibinfo {year} {2023})},\ \Eprint
  {http://arxiv.org/abs/2307.14416} {arXiv:2307.14416 [hep-th]} \BibitemShut
  {NoStop}%
\bibitem [{\citenamefont {Antonini}\ and\ \citenamefont
  {Rath}(2024)}]{AntRat24}%
  \BibitemOpen
  \bibfield  {author} {\bibinfo {author} {\bibfnamefont {Stefano}\ \bibnamefont
  {Antonini}}\ and\ \bibinfo {author} {\bibfnamefont {Pratik}\ \bibnamefont
  {Rath}},\ }\bibfield  {title} {\enquote {\bibinfo {title} {{Do holographic
  CFT states have unique semiclassical bulk duals?}}}\ }\href@noop {} {\
  (\bibinfo {year} {2024})},\ \Eprint {http://arxiv.org/abs/2408.02720}
  {arXiv:2408.02720 [hep-th]} \BibitemShut {NoStop}%
\bibitem [{\citenamefont {Hamilton}\ \emph {et~al.}(2006)\citenamefont
  {Hamilton}, \citenamefont {Kabat}, \citenamefont {Lifschytz},\ and\
  \citenamefont {Lowe}}]{HamKab05}%
  \BibitemOpen
  \bibfield  {author} {\bibinfo {author} {\bibfnamefont {Alex}\ \bibnamefont
  {Hamilton}}, \bibinfo {author} {\bibfnamefont {Daniel~N.}\ \bibnamefont
  {Kabat}}, \bibinfo {author} {\bibfnamefont {Gilad}\ \bibnamefont
  {Lifschytz}}, \ and\ \bibinfo {author} {\bibfnamefont {David~A.}\
  \bibnamefont {Lowe}},\ }\bibfield  {title} {\enquote {\bibinfo {title}
  {{Local bulk operators in AdS/CFT: A Boundary view of horizons and
  locality}},}\ }\href {\doibase 10.1103/PhysRevD.73.086003} {\bibfield
  {journal} {\bibinfo  {journal} {Phys.Rev.}\ }\textbf {\bibinfo {volume}
  {D73}},\ \bibinfo {pages} {086003} (\bibinfo {year} {2006})},\ \Eprint
  {http://arxiv.org/abs/hep-th/0506118} {arXiv:hep-th/0506118 [hep-th]}
  \BibitemShut {NoStop}%
\bibitem [{\citenamefont {Harlow}\ and\ \citenamefont
  {Numasawa}(2023)}]{Harlow:2023hjb}%
  \BibitemOpen
  \bibfield  {author} {\bibinfo {author} {\bibfnamefont {Daniel}\ \bibnamefont
  {Harlow}}\ and\ \bibinfo {author} {\bibfnamefont {Tokiro}\ \bibnamefont
  {Numasawa}},\ }\bibfield  {title} {\enquote {\bibinfo {title} {{Gauging
  spacetime inversions in quantum gravity}},}\ }\href@noop {} {\  (\bibinfo
  {year} {2023})},\ \Eprint {http://arxiv.org/abs/2311.09978} {arXiv:2311.09978
  [hep-th]} \BibitemShut {NoStop}%
\bibitem [{\citenamefont {Brown}\ \emph {et~al.}(2020)\citenamefont {Brown},
  \citenamefont {Gharibyan}, \citenamefont {Penington},\ and\ \citenamefont
  {Susskind}}]{BroGha19}%
  \BibitemOpen
  \bibfield  {author} {\bibinfo {author} {\bibfnamefont {Adam~R.}\ \bibnamefont
  {Brown}}, \bibinfo {author} {\bibfnamefont {Hrant}\ \bibnamefont
  {Gharibyan}}, \bibinfo {author} {\bibfnamefont {Geoff}\ \bibnamefont
  {Penington}}, \ and\ \bibinfo {author} {\bibfnamefont {Leonard}\ \bibnamefont
  {Susskind}},\ }\bibfield  {title} {\enquote {\bibinfo {title} {{The
  Python\textquoteright{}s Lunch: geometric obstructions to decoding Hawking
  radiation}},}\ }\href {\doibase 10.1007/JHEP08(2020)121} {\bibfield
  {journal} {\bibinfo  {journal} {JHEP}\ }\textbf {\bibinfo {volume} {08}},\
  \bibinfo {pages} {121} (\bibinfo {year} {2020})},\ \Eprint
  {http://arxiv.org/abs/1912.00228} {arXiv:1912.00228 [hep-th]} \BibitemShut
  {NoStop}%
\bibitem [{\citenamefont {Engelhardt}\ \emph {et~al.}(2022)\citenamefont
  {Engelhardt}, \citenamefont {Penington},\ and\ \citenamefont
  {Shahbazi-Moghaddam}}]{EngPen21b}%
  \BibitemOpen
  \bibfield  {author} {\bibinfo {author} {\bibfnamefont {Netta}\ \bibnamefont
  {Engelhardt}}, \bibinfo {author} {\bibfnamefont {Geoff}\ \bibnamefont
  {Penington}}, \ and\ \bibinfo {author} {\bibfnamefont {Arvin}\ \bibnamefont
  {Shahbazi-Moghaddam}},\ }\bibfield  {title} {\enquote {\bibinfo {title}
  {{Finding pythons in unexpected places}},}\ }\href {\doibase
  10.1088/1361-6382/ac3e75} {\bibfield  {journal} {\bibinfo  {journal} {Class.
  Quant. Grav.}\ }\textbf {\bibinfo {volume} {39}},\ \bibinfo {pages} {094002}
  (\bibinfo {year} {2022})},\ \Eprint {http://arxiv.org/abs/2105.09316}
  {arXiv:2105.09316 [hep-th]} \BibitemShut {NoStop}%
\bibitem [{\citenamefont {Heemskerk}\ \emph {et~al.}(2012)\citenamefont
  {Heemskerk}, \citenamefont {Marolf}, \citenamefont {Polchinski},\ and\
  \citenamefont {Sully}}]{HeeMar12}%
  \BibitemOpen
  \bibfield  {author} {\bibinfo {author} {\bibfnamefont {Idse}\ \bibnamefont
  {Heemskerk}}, \bibinfo {author} {\bibfnamefont {Donald}\ \bibnamefont
  {Marolf}}, \bibinfo {author} {\bibfnamefont {Joseph}\ \bibnamefont
  {Polchinski}}, \ and\ \bibinfo {author} {\bibfnamefont {James}\ \bibnamefont
  {Sully}},\ }\bibfield  {title} {\enquote {\bibinfo {title} {{Bulk and
  Transhorizon Measurements in AdS/CFT}},}\ }\href {\doibase
  10.1007/JHEP10(2012)165} {\bibfield  {journal} {\bibinfo  {journal} {JHEP}\
  }\textbf {\bibinfo {volume} {10}},\ \bibinfo {pages} {165} (\bibinfo {year}
  {2012})},\ \Eprint {http://arxiv.org/abs/1201.3664} {arXiv:1201.3664
  [hep-th]} \BibitemShut {NoStop}%
\end{thebibliography}%

\end{document}